\documentclass[aps,pra,reprint,showpacs,superscriptaddress,twocolumn,notitlepage,10pt]{revtex4-1}%
\usepackage{amssymb}
\usepackage{mathrsfs}
\usepackage{amsfonts}
\usepackage{graphicx}
\usepackage{amsmath}
\usepackage{dcolumn}
\usepackage{color}
\usepackage[colorlinks,linkcolor=blue,citecolor=blue,hyperindex,bookmarks=false,pdfstartview=FitH]{hyperref}
\newcommand{\red}[1]{\textcolor[rgb]{0.80,0.00,0.00}{#1}}

\newcommand{\figpanel}[2]{\hyperref[#1]{\ref*{#1}(#2)}}

\begin{document}
\title{Nonreciprocal frequency conversion with chiral $\Lambda$-type atoms}
\author{Lei Du}
\affiliation{Beijing Computational Science Research Center, Beijing 100193, China}
\author{Yao-Tong Chen}
\affiliation{Center for Quantum Sciences and School of Physics, Northeast Normal University, Changchun 130117, China}
\author{Yong Li}
\email{liyong@csrc.ac.cn}
\affiliation{Beijing Computational Science Research Center, Beijing 100193, China}

\date{\today}

\begin{abstract}
In this paper, we begin with a model of a $\Lambda$-type atom whose both transitions are chirally coupled to a waveguide and then extend the model to its giant-atom version. We investigate the single-photon scatterings of the giant-atom model in both the Markovian and non-Markovian regimes. It is shown that the chiral atom-waveguide couplings enable nonreciprocal, reflectionless, and efficient frequency conversion, while the giant-atom structure introduces intriguing interference effects to the scattering behaviors, such as ultra-narrow scattering windows. The chiral giant-atom model exhibits quite different scattering spectra in the two regimes and, in particular, demonstrates non-Markovicity induced nonreciprocity under specific conditions. These phenomena can be understood from the effective detuning and decay rate of the giant-atom model. We believe that our results have potential applications in integrated photonics and quantum network engineering.   
\end{abstract}

\maketitle

\section{Introduction}

Nonreciprocal photon transmissions play significant roles in fabricating various integrated photonic elements and thus have important applications ranging from quantum information processing to quantum network engineering~\cite{NR1,NR2,NR3}. In particular, it is beneficial to break the Lorentz reciprocity at the single-photon level to avoid the harmful backflows between quantum devices and thereby to protect the signal sources. In general, nonreciprocal photon transmissions can be achieved via magneto-optical effects~\cite{magop1,magop2,magop3,magop4,magop5}, optical nonlinearities~\cite{opNL1,opNL2,opNL3,opNL4}, dynamic modulations~\cite{dymod1,dymod2,dymod3,dymod4,dymod5,dymod6,dymod7,dymod8,dymod9,dymod10,dymod11}, synthetic magnetism~\cite{syn1,syn2,syn3,syn4,syn5,syn6,syn7}, and reservoir engineerings~\cite{clerkX,clerkNP}.

On the other hand, the interactions between light and matter (e.g., atoms) can also be direction dependent: atoms can interact differently with photons propagating along different directions. Such phenomena, known as ``chiral quantum optics'' in the literature~\cite{chiralZoller,chiralAW}, provide a new paradigm for quantum network engineering. To date, significant progress has been made on the basis of chiral quantum optics, such as cascaded quantum systems~\cite{cas1,cas2,cas3,cas4}, deterministic photon routing~\cite{rout1,rout2}, and non-destructive photon detection~\cite{detec}. Moreover, non-Markovian retardation effects, e.g., time-delayed quantum feedbacks, have also been widely investigated in chiral quantum optical systems~\cite{FB1,FB2,FB3,FB4,FB5}. In experiments, chiral atom-field interactions can be achieved via several approaches, such as the spin-momentum locking effect of light in one-dimensional optical fibers~\cite{chiralfiber1,chiralfiber2,chiralfiber3}, inserting circulators in superconducting circuits~\cite{SCchiral1,SCchiral2,SCchiral3}, topological engineering~\cite{topo1,topo2}, and synthesizing artificial gauge fields~\cite{sawtooth}. In spite of the broken time-reversal symmetry, however, it has been shown that such chirality is not sufficient to support nonreciprocal scatterings solely in the absence of intrinsic atomic dissipations~\cite{yanweibin}. To achieve high-performance nonreciprocal transmissions, one should exploit, for instance, considerable intrinsic atomic dissipations or atoms with more transitions chirally coupled to the waveguide field~\cite{rout1,shenchiral,extract}.

In this paper, we begin with the small-atom model considered in Ref.~\cite{rout1}, where both the transitions of a $\Lambda$-type atom are chirally coupled to the waveguide at a single point. We briefly review the nonreciprocal single-photon scatterings of this model, which are allowed even if the intrinsic atomic dissipation is ignored. This is quite different from the case of a chiral two-level atom, where the intrinsic dissipations are indispensable for observing nonreciprocal scatterings. Afterwards, we extend this chiral $\Lambda$-type atom to its giant-atom version, where both the atomic transitions couple chirally to the waveguide at two separated points with the separation distance comparable to the wavelength of the waveguide field. In the past few years, nonchiral giant atoms have been well investigated~\cite{fiveyears}, which demonstrated a series of intriguing interference effects such as frequency-dependent Lamb shift and relaxation rate~\cite{Lamb}, decoherence-free interatomic interactions~\cite{decofree1,decofree2,decofree3}, various bound states~\cite{oscillate,WZHbound,GAYuan}, and modified topological effects~\cite{GACheng,GATudela}. Most recently, the concept of chiral quantum optics has also been brought to giant-atom structures~\cite{WXprl,AFKchiral,WXarxiv,DLsyn}, which shows the possibility of combining the advantages of the two paradigms. Here, we study the nonreciprocal single-photon frequency conversion of the chiral $\Lambda$-type giant atom in both the Markovian and non-Markovian regimes, which are defined depending on whether the propagation time of photons between different atom-waveguide coupling points is negligible or not. In both regimes, the chiral giant-atom model exhibits intriguing interference effects such as ultra-narrow scattering windows. In particular, non-Markovicity induced nonreciprocal scatterings are demonstrated under specific conditions.             

\section{Chiral small atom}

\begin{figure}[ptb]
\centering
\includegraphics[width=8 cm]{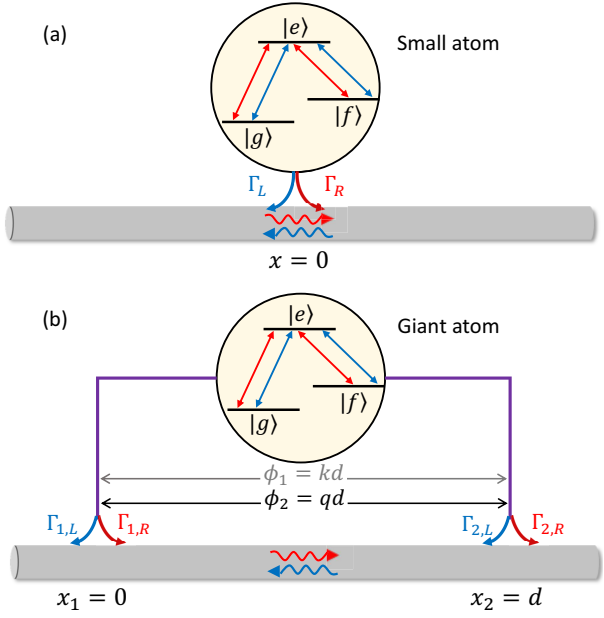}
\caption{Schematic diagram of (a) the $\Lambda$-type small atom with chiral atom-waveguide couplings at $x=0$ and (b) the $\Lambda$-type giant atom with chiral atom-waveguide couplings at both coupling points $x_{1}=0$ and $x_{2}=d$.}\label{fig1}
\end{figure}

We begin by considering a $\Lambda$-type small atom, where both the atomic transitions ($|g\rangle\leftrightarrow|e\rangle$ and $|f\rangle\leftrightarrow|e\rangle$) interact chirally with the waveguide field at $x=0$, as shown in Fig.~\figpanel{fig1}{a}. Similar model has been considered in Ref.~\cite{rout1} to demonstrate nonreciprocal few-photon scatterings. However, we briefly review this model here to pave the way for introducing the giant-atom model in the next section and facilitate the comparison between the two models. The Hamiltonian of such a small-atom model can be written as ($\hbar=1$ hereafter)
\begin{equation}
\begin{split}
H&=H_{\text{a}}+H_{\text{w}}+H_{\text{int}},\\
H_{\text{a}}&=\omega_{f}|f\rangle\langle f|+\omega_{e}|e\rangle\langle e|,\\
H_{\text{w}}&=\int_{-\infty}^{+\infty}dx\Big[c_{L}^{\dag}(x)\Big(\omega_{0}+iv_{g}\frac{\partial}{\partial x}\Big)c_{L}(x)\\
&\quad\,+c_{R}^{\dag}(x)\Big(\omega_{0}-iv_{g}\frac{\partial}{\partial x}\Big)c_{R}(x)\Big],\\
H_{\text{int}}&=\int_{-\infty}^{\infty}dx \delta(x)\Big\{|g\rangle\langle e|[g_{R}c_{R}^{\dag}(x)+g_{L}c_{L}^{\dag}(x)]\\
&\quad+|f\rangle\langle e|[\xi_{R}c_{R}^{\dag}(x)+\xi_{L}c_{L}^{\dag}(x)]+\text{H.c.}\Big\},
\end{split}
\label{eq1}
\end{equation}
where $H_{\text{a}}$ is the free Hamiltonian of the atom, with $\omega_{f}$ and $\omega_{e}$ the energies of the states $|f\rangle$ and $|e\rangle$ with respect to the state $|g\rangle$, respectively; $H_{\text{w}}$ is the Hamiltonian of the waveguide, with $c_{R}^{\dag}(x)$ [$c_{L}^{\dag}(x)$] the creation operator of the right-moving (left-moving) waveguide mode at position $x$ and $v_{g}$ the group velocity; $\omega_{0}$ is a chosen frequency that is away from the cutoff frequency, around which the dispersion relation of the waveguide can be linearized as $E=\omega_{0}+kv_{g}$ ($k$ is the renormalized wave vector of photons in the waveguide)~\cite{shen2005,shen2009}; $H_{\text{int}}$ describes the interactions between the atomic transitions and the waveguide modes, with $g_{R}$ and $g_{L}$ ($\xi_{R}$ and $\xi_{L}$) the coupling strengths of the transition $|g\rangle\leftrightarrow|e\rangle$ ($|f\rangle\leftrightarrow|e\rangle$) with the right-moving and left-moving waveguide modes (assumed to be real), respectively. For simplicity, we assume $\xi_{R}=g_{R}$ and $\xi_{L}=g_{L}$ in this section. Moreover, we have assumed that the two lower states $|g\rangle$ and $|f\rangle$ are quasi-degenerate, such that both the transition frequencies $\omega_{e}$ and $\omega_{e}-\omega_{f}$ fall into the linearized region around $\omega_{0}$~\cite{NJPLam,shenLam,DLlambda,FanLambda}. The case of nondegenerate lower states will be discussed in Sec.~\ref{s3d}, where the group velocities corresponding to the two transition frequencies are different, yet our main results are shown to be insensitive to such a difference.    

In the single-excitation space, the eigenstate of Hamiltonian (\ref{eq1}) can be written as
\begin{equation}
\begin{split}
|\psi\rangle&=\sum_{\alpha=g,f}\int_{-\infty}^{+\infty}dx\Big[R_{\alpha}(x)c_{R}^{\dag}(x)\\
&\quad\,+L_{\alpha}(x)c_{L}^{\dag}(x)\Big]|0,\alpha\rangle+u_{e}|0,e\rangle,
\end{split}
\label{eq2}
\end{equation}
where $R_{\alpha}(x)$ [$L_{\alpha}(x)$] is the probability amplitude of finding a right-moving (left-moving) photon in the waveguide at position $x$ and the atom finally in the state $|\alpha\rangle$; $u_{e}$ is the probability amplitude of the atom in the excited state $|e\rangle$. Solving the stationary Schr\"{o}dinger equation $H|\psi\rangle=E|\psi\rangle$ with Eqs.~(\ref{eq1}) and (\ref{eq2}), one can obtain
\begin{equation}
\begin{split}
ER_{g}(x)&=\Big(\omega_{0}-iv_{g}\frac{\partial}{\partial x}\Big)R_{g}(x)+g_{R}\delta(x)u_{e},\\
EL_{g}(x)&=\Big(\omega_{0}+iv_{g}\frac{\partial}{\partial x}\Big)L_{g}(x)+g_{L}\delta(x)u_{e},\\
ER_{f}(x)&=\Big(\omega_{0}+\omega_{f}-iv_{g}\frac{\partial}{\partial x}\Big)R_{f}(x)+g_{R}\delta(x)u_{e},\\
EL_{f}(x)&=\Big(\omega_{0}+\omega_{f}+iv_{g}\frac{\partial}{\partial x}\Big)L_{f}(x)+g_{L}\delta(x)u_{e},\\
\Delta u_{e}&=g_{R}[R_{g}(0)+R_{f}(0)]+g_{L}[L_{g}(0)+L_{f}(0)],
\end{split}
\label{eq3}
\end{equation}
where $\Delta=E-\omega_{e}$ is the detuning between the propagating photons and the transition $|g\rangle\leftrightarrow|e\rangle$. 

We first consider that a single photon is incident from the left side of the waveguide and the atom occupies the state $|g\rangle$ initially. If the atom is excited by the photon, then the atom can spontaneously decay back to $|g\rangle$ and reemit a photon with the same wave vector $k$, or it can decay to another lower state $|f\rangle$ and reemit a photon with wave vector $q=k-\omega_{f}/v_{g}$~\cite{NJPLam,shenLam,DLlambda,FanLambda}. We refer to these two processes as elastic and inelastic scatterings, respectively, depending on whether the frequency of the input photon is converted or not. For the left-incident photon here, one can assume
\begin{equation}
\begin{split}
R_{g}(x)&=[\Theta(-x)+t_{1}\Theta(x)]e^{ikx},\\
L_{g}(x)&=r_{1}\Theta(-x)e^{-ikx},\\
R_{f}(x)&=t_{2}\Theta(x)e^{iqx},\\
L_{f}(x)&=r_{2}\Theta(-x)e^{-iqx},
\end{split}
\label{eq4}
\end{equation}
where $\Theta(x)$ is the Heaviside step function; $t_{1}$ and $r_{1}$ ($t_{2}$ and $r_{2}$) are the transmission and reflection coefficients of the elastic (inelastic) scattering processes, respectively. Substituting Eq.~(\ref{eq4}) into Eq.~(\ref{eq3}), one readily obtains  
\begin{equation}
\begin{split}
t_{1}&=\frac{\Delta+i\Gamma_{L}}{\Delta+i(\Gamma_{L}+\Gamma_{R})},\\
t_{2}&=\frac{-i\Gamma_{R}}{\Delta+i(\Gamma_{R}+\Gamma_{L})},
\end{split}
\label{eq5}
\end{equation} 
where $\Gamma_{R}=g_{R}^{2}/v_{g}$ and $\Gamma_{L}=g_{L}^{2}/v_{g}$ are the radiative decay rates of $|e\rangle$ induced by the right-moving and left-moving waveguide modes, respectively. Note that we only focus on the two transmission coefficients in this paper, with which it is enough to understand the underlying physics of our models.

Similarly, for a single photon incident from the right side of the waveguide (the atom is still assumed in the state $|g\rangle$ initially), with the ansatz
\begin{equation}
\begin{split}
R_{g}(x)&=\tilde{r}_{1}\Theta(x)e^{ikx},\\
L_{g}(x)&=[\Theta(x)+\tilde{t}_{1}\Theta(-x)]e^{-ikx},\\
R_{f}(x)&=\tilde{r}_{2}\Theta(x)e^{iqx},\\
L_{f}(x)&=\tilde{t}_{2}\Theta(-x)e^{-iqx},
\end{split}
\label{eq6}
\end{equation}
the elastic and inelastic transmission coefficients in this case can be solved as
\begin{equation}
\begin{split}
\tilde{t}_{1}&=\frac{\Delta+i\Gamma_{R}}{\Delta+i(\Gamma_{R}+\Gamma_{L})},\\
\tilde{t}_{2}&=\frac{-i\Gamma_{L}}{\Delta+i(\Gamma_{R}+\Gamma_{L})}.
\end{split}
\label{eq7}
\end{equation}

\begin{figure}[ptb]
\centering
\includegraphics[width=8.5 cm]{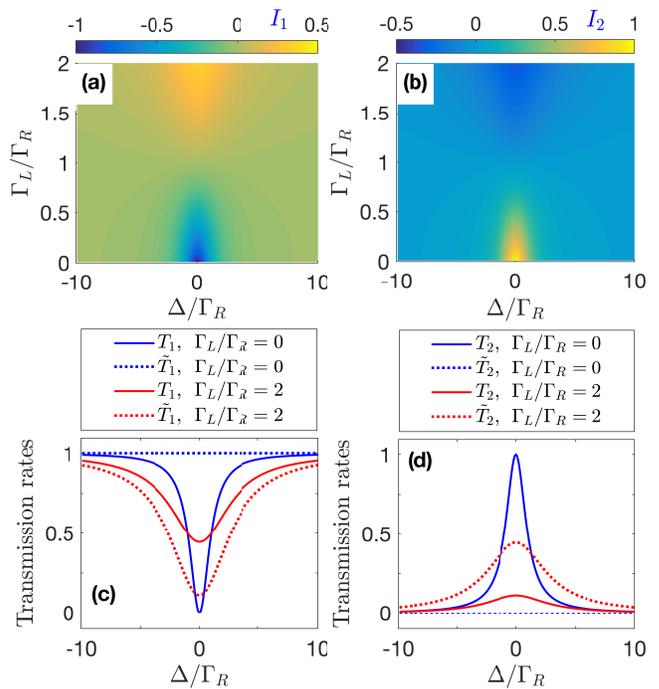}
\caption{Psuedocolormaps of transmission contrasts (a) $I_{1}$ and (b) $I_{2}$ of the chiral small-atom model. Profiles of (c) elastic and (d) inelastic transmission rates with different values of $\Gamma_{L}/\Gamma_{R}$.}\label{fig2}
\end{figure}

It is clear from Eqs.~(\ref{eq5}) and (\ref{eq7}) that nonreciprocal scatterings (i.e., $|t_{1}|\neq|\tilde{t}_{1}|$ and/or $|t_{2}|\neq|\tilde{t}_{2}|$) can be achieved if $g_{R}\neq g_{L}$ (i.e., $\Gamma_{R}\neq\Gamma_{L}$), yet this is impossible for the case of a chiral two-level atom in the absence of intrinsic atomic dissipations (see Appendix~\ref{app1} for more details). Physically, the nonreciprocity here is because, for an input photon that can be absorbed by the transition $|g\rangle\leftrightarrow|e\rangle$, the other transition $|f\rangle\leftrightarrow|e\rangle$ effectively serves as a ``dissipation channel'', which makes the elastic scattering process behaves like that of a two-level atom with intrinsic dissipations. The excitation probabilities of the atom are different for photons propagating along opposite directions due to the chiral atom-waveguide couplings, which thus result in nonreciprocal elastic and inelastic scatterings. 

Figures~\figpanel{fig2}{a} and \figpanel{fig2}{b} depict the transmission contrasts $I_{1}=T_{1}-\tilde{T}_{1}$ and $I_{2}=T_{2}-\tilde{T}_{2}$ versus the detuning $\Delta$ and the decay ratio $\Gamma_{L}/\Gamma_{R}$, where $T_{1}=|t_{1}|^{2}$ and $\tilde{T}_{1}=|\tilde{t}_{1}|^{2}$ ($T_{2}=|t_{2}|^{2}$ and $\tilde{T}_{2}=|\tilde{t}_{2}|^{2}$) are the elastic (inelastic) transmission rates for the left-incident and right-incident photons, respectively. It can be seen that the optimal nonreciprocal scatterings ($|I_{1}|=|I_{2}|=1$) occur if $\Gamma_{L}=0$ (or $\Gamma_{R}=0$, leading to reversed nonreciprocal scatterings), while the scatterings are totally reciprocal ($I_{1}=I_{2}=0$) if $\Gamma_{L}=\Gamma_{R}$. To show more details, we also plot in Figs.~\figpanel{fig2}{c} and \figpanel{fig2}{d} the profiles of the transmission rates of two cases, i.e., the ideal chiral case with $\Gamma_{L}/\Gamma_{R}=0$ and the nonideal chiral case with $\Gamma_{L}/\Gamma_{R}=2$. In the ideal chiral case, the elastic (inelastic) transmission rate of the left-incident single photon shows a standard anti-Lorentzian (Lorentzian) line shape and efficient frequency conversion with $T_{2}=1$ can be achieved, whereas for a right-incident photon, the inelastic scattering is totally suppressed ($\tilde{T}_{2}\equiv0$) because the atom is decoupled with the left-moving waveguide mode. In the nonideal chiral case, the atom can still exhibit nonreciprocal transmissions, yet both the transmission contrasts and the conversion efficiency are degraded in this case. In view of this, we focus on the ideal chiral case hereafter in this paper. 

\section{Chiral giant atom}

Now we consider a giant-atom version of the chiral $\Lambda$-type atom, where both the transitions $|g\rangle\leftrightarrow|e\rangle$ and $|f\rangle\leftrightarrow|e\rangle$ are coupled chirally with the waveguide modes at two separate points $x_{1}=0$ and $x_{2}=d$, as shown in Fig.~\figpanel{fig1}{b}. The case of a nonchiral $\Lambda$-type giant atom has been investigated in our previous work~\cite{DLlambda}, where the optimal frequency conversion efficiency is still at most one half in spite of the giant-atom interference effects. In this section, we would like to demonstrate that while the chiral couplings enable nonreciprocal and efficient frequency conversion as shown above, the giant-atom structure brings some intriguing interference effects, especially in the non-Markovian regime.

For the giant atom considered here, the interaction Hamiltonian becomes
\begin{equation}
\begin{split}
H_{\text{int}}&=\sum_{\alpha=g,f}\int_{-\infty}^{\infty}dx |\alpha\rangle\langle e|\Big\{c_{R}^{\dag}(x)[g_{1,R}\delta(x)\\
&\quad\,+g_{2,R}\delta(x-d)]+c_{L}^{\dag}(x)[g_{1,L}\delta(x)\\
&\quad\,+g_{2,L}\delta(x-d)]\Big\}+\text{H.c.}.
\end{split}
\label{eq8}
\end{equation}  
Here, $g_{1,R}$ and $g_{2,R}$ ($g_{1,L}$ and $g_{2,L}$) are the coupling strengths (assumed to be real again) between the atomic transitions and the right-moving (left-moving) waveguide modes at the points $x=x_{1}$ and $x=x_{2}$, respectively. Similar to the small-atom case, we have assumed that both the atomic transitions couple to the right-moving (left-moving) waveguide mode with the same strength $g_{j,R}$ ($g_{j,L}$) at each coupling point ($j=1,2$). In this case, the probability amplitudes of the waveguide modes can be written as
\begin{equation}
\begin{split}
R_{g}(x)&=\{\Theta(-x)+A[\Theta(x)-\Theta(x-d)]\\
&\quad\,+t_{1}\Theta(x-d)\}e^{ikx},\\
L_{g}(x)&=\{r_{1}\Theta(-x)+B[\Theta(x)-\Theta(x-d)]\}e^{-ikx},\\
R_{f}(x)&=\{M[\Theta(x)-\Theta(x-d)]+{t}_{2}\Theta(x-d)\}e^{iqx},\\
L_{f}(x)&=\{r_{2}\Theta(-x)+N[\Theta(x)-\Theta(x-d)]\}e^{-iqx}
\end{split}
\label{eq9}
\end{equation}
for a left-incident photon and 
\begin{equation}
\begin{split}
R_{g}(x)&=\{\tilde{r}_{1}\Theta(x-d)+\tilde{A}[\Theta(x)-\Theta(x-d)]\}e^{ikx},\\
L_{g}(x)&=\{\Theta(x-d)+\tilde{B}[\Theta(x)-\Theta(x-d)]\\
&\quad\,+\tilde{t}_{1}\Theta(-x)\}e^{-ikx},\\
R_{f}(x)&=\{\tilde{r}_{2}\Theta(x-d)+\tilde{M}[\Theta(x)-\Theta(x-d)]\}e^{iqx},\\
L_{f}(x)&=\{\tilde{N}[\Theta(x)-\Theta(x-d)]+\tilde{t}_{2}\Theta(-x)\}e^{-iqx}
\end{split}
\label{eq10}
\end{equation}
for a right-incident photon. Here $A$ and $B$ ($M$ and $N$) are the probability amplitudes of finding right-moving and left-moving photons with wave vector $k$ ($q$) in the region of $x_{1}<x<x_{2}$, respectively, if the photon is incident from the left side, and those with tildes correspond to a right-incident photon. Then the transmission coefficients can be solved as
\begin{equation}
\begin{split}
t_{1}&=\frac{\Delta+i(\Gamma_{1,L}+\Gamma_{2,L}+F_{-})}{\Delta+i(\Gamma_{1,R}+\Gamma_{2,R}+\Gamma_{1,L}+\Gamma_{2,L}+F_{+})},\\
t_{2}&=\frac{-i[\Gamma_{1,R}+\Gamma_{2,R}e^{i(\phi_{1}-\phi_{2})}+\Gamma_{12,R}(e^{i\phi_{1}}+e^{-i\phi_{2}})]}{\Delta+i(\Gamma_{1,R}+\Gamma_{2,R}+\Gamma_{1,L}+\Gamma_{2,L}+F_{+})},\\
\tilde{t}_{1}&=\frac{\Delta+i(\Gamma_{1,R}+\Gamma_{2,R}+\tilde{F}_{-})}{\Delta+i(\Gamma_{1,R}+\Gamma_{2,R}+\Gamma_{1,L}+\Gamma_{2,L}+F_{+})},\\
\tilde{t}_{2}&=\frac{-i[\Gamma_{1,L}+\Gamma_{2,L}e^{i(\phi_{2}-\phi_{1})}+\Gamma_{12,L}(e^{-i\phi_{1}}+e^{i\phi_{2}})]}{\Delta+i(\Gamma_{1,R}+\Gamma_{2,R}+\Gamma_{1,L}+\Gamma_{2,L}+F_{+})},
\end{split}
\label{eq11}
\end{equation}
with
\begin{equation}
\begin{split}
F_{+}&=(\Gamma_{12,R}+\Gamma_{12,L})(e^{i\phi_{1}}+e^{i\phi_{2}}),\\
F_{-}&=\Gamma_{12,R}(e^{i\phi_{2}}-e^{-i\phi_{1}})+\Gamma_{12,L}(e^{i\phi_{1}}+e^{i\phi_{2}}),\\
\tilde{F}_{-}&=\Gamma_{12,R}(e^{i\phi_{1}}+e^{i\phi_{2}})+\Gamma_{12,L}(e^{i\phi_{2}}-e^{-i\phi_{1}}),
\end{split}
\label{eq12}
\end{equation}
where $\Gamma_{j,\beta}=g_{j,\beta}^{2}/v_{g}$ and $\Gamma_{12,\beta}=\sqrt{\Gamma_{1,\beta}\Gamma_{2,\beta}}$ ($\beta=L,R$); $\phi_{1}=kd=(\omega_{e}-\omega_{0}+\Delta)\tau$ and $\phi_{2}=qd=(\omega_{e}-\omega_{f}-\omega_{0}+\Delta)\tau$ are the phases of photons accumulated between the two atom-waveguide coupling points, with $\tau=d/v_{g}$ the corresponding propagation time. Clearly, both $\phi_{1}$ and $\phi_{2}$ consist of a constant part and a $\Delta$-dependent part, such that we define $\phi_{1}=\phi_{1,0}+\tau\Delta$ and $\phi_{2}=\phi_{2,0}+\tau\Delta$ for simplicity with $\phi_{1,0}=(\omega_{e}-\omega_{0})\tau$ and $\phi_{2,0}=(\omega_{e}-\omega_{f}-\omega_{0})\tau$. In this way, we can study single-photon scatterings in both the Markovian and non-Markovian regimes, depending on whether the propagation time $\tau$ is negligible or not, as discussed in detail below. Note that we focus on the ideal chiral case of $\Gamma_{1,L}=\Gamma_{2,L}=0$ in this section, which enables efficient frequency conversion as discussed above.

\begin{figure}[ptb]
\centering
\includegraphics[width=8.5 cm]{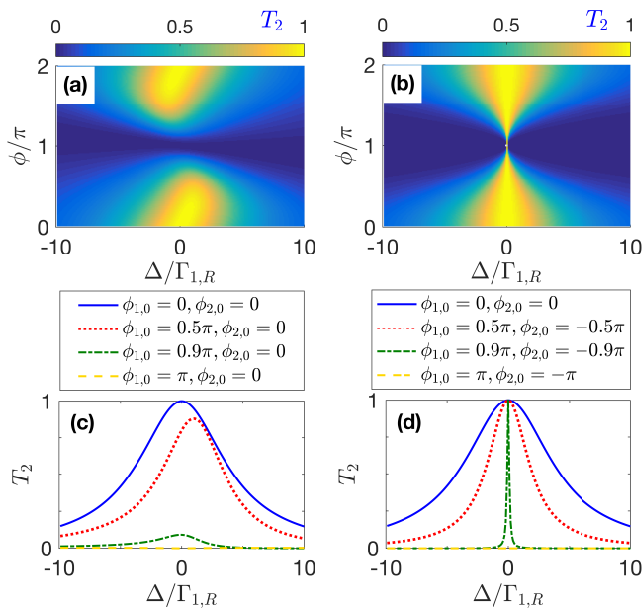}
\caption{Psuedocolormaps of inelastic transmission rate $T_{2}$ of the chiral giant-atom model in the Markovian regime with (a) $\phi_{1,0}=\phi$, $\phi_{2,0}=0$ and (b) $\phi_{1,0}=-\phi_{2,0}=\phi$. (c) and (d) depict the profiles of $T_{2}$ with different values of phases, which correspond to the cases of (a) and (b), respectively. Here we assume $\Gamma_{1,R}=\Gamma_{2,R}$, $\Gamma_{1,L}=\Gamma_{2,L}=0$, and $\tau\Gamma_{1,R}=0.03$.}\label{fig3}
\end{figure}

\subsection{Markovian regime}

We first consider the Markovian regime, where the propagation time $\tau$ is small enough ($\tau\sum_{j,\beta}\Gamma_{j,\beta}\ll 1$~\cite{GuoSAW,longhiretard}) such that $\phi_{1}\approx\phi_{1,0}$ and $\phi_{2}\approx\phi_{2,0}$ are approximately constant: for small enough $\tau$, $\text{exp}(i\phi_{1})\approx\text{exp}(i\phi_{1,0})$ and $\text{exp}(i\phi_{2})\approx\text{exp}(i\phi_{2,0})$ can be obtained with Taylor expansion to the first order of $\tau$. 

Figures~\figpanel{fig3}{a} and \figpanel{fig3}{b} depict the inelastic transmission rate $T_{2}$ of a left-incident photon versus the detuning $\Delta$ and the phase factor $\phi$ [we assume $\phi_{1,0}=\phi$, $\phi_{2,0}=0$ in Fig.~\figpanel{fig3}{a} and $\phi_{1,0}=-\phi_{2,0}=\phi$ in Fig.~\figpanel{fig3}{b}] in the ideal chiral case. In this case, the elastic transmission rate $T_{1}$ can be readily obtained with $T_{1}=1-T_{2}$, while the transmission rates of a right-incident photon are $\tilde{T}_{1}\equiv1$ and $\tilde{T}_{2}\equiv0$. One can find that the pattern in Fig.~\figpanel{fig3}{a} is quite similar to that of a nonchiral $\Lambda$-type giant atom~\cite{DLlambda}: (i) the scattering behaviors are phase dependent with the periodicity of $2\pi$; (ii) the frequency conversion is completely suppressed (i.e., $T_{2}\equiv0$) if $\phi_{1,0}=(2m+1)\pi$ ($m$ is an arbitrary integer) because the transition $|g\rangle\leftrightarrow|e\rangle$ is decoupled from the waveguide modes in this case. Note that the frequency conversion is also unachievable when $\phi_{2,0}=(2m+1)\pi$ (not shown here). However, perfect elastic transmission (i.e., $T_{1}\equiv1$) takes place in this case due to the chiral atom-waveguide couplings, which is different from the nonchiral model that shows total reflection~\cite{DLlambda}. 

More interestingly, efficient frequency conversion with ultra-narrow window can be achieved if $\phi_{1,0}=-\phi_{2,0}=\phi$ and $\phi\rightarrow\pi$. As shown in Fig.~\figpanel{fig3}{b}, the width of the conversion window decreases gradually as $\phi$ increases from $0$ to $\pi$, yet the peak value of $T_{2}$ keeps invariant during this process. However, the scattering behaviors change abruptly at $\phi=\pi$ where the frequency conversion is completely suppressed, as shown analytically in Eq.~(\ref{eq11}). Then the conversion window reopens with its width increasing with $\phi$ and the peak value keeping invariant again. Such ultra-narrow scattering windows are supposed to have applications in, e.g., precise frequency conversion and sensing, and can be understood from the effective decay rate of the atom, as discussed in detail in Appendix~\ref{appb}.

The results above can be seen in a clearer way via the two-dimensional profiles in Figs.~\figpanel{fig3}{c} and \figpanel{fig3}{d}. For $\phi_{1,0}=\phi$ and $\phi_{2,0}=0$, the maximal conversion efficiency decreases gradually as $\phi$ increases from $0$ to $\pi$, although the width of the window can still be suppressed. The position of the conversion peak is also phase dependent, similar to that of the nonchiral case, yet the profile becomes slightly asymmetric if $\phi\neq m\pi$ due to the finite value of $\tau$. For $\phi_{1,0}=-\phi_{2,0}=\phi$, however, both the peak position and the maximal conversion efficiency keep invariant, while the linewidth decreases markedly as $\phi$ increases from $0$ until the conversion window disappears abruptly at $\phi=\pi$ as shown above.   

\subsection{Non-Markovian regime}

\begin{figure}[ptb]
\centering
\includegraphics[width=8.5 cm]{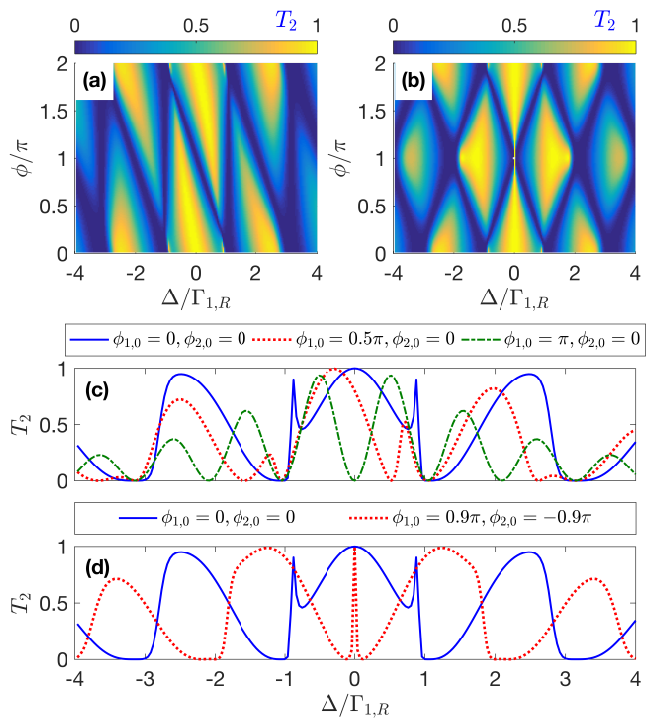}
\caption{Psuedocolormaps of inelastic transmission rate $T_{2}$ of the chiral giant-atom model in the non-Markovian regime with (a) $\phi_{1,0}=\phi$, $\phi_{2,0}=0$ and (b) $\phi_{1,0}=-\phi_{2,0}=\phi$. (c) and (d) depict the profiles of $T_{2}$ with different values of phases, which correspond to the cases of (a) and (b), respectively. Here we assume $\Gamma_{1,R}=\Gamma_{2,R}$, $\Gamma_{1,L}=\Gamma_{2,L}=0$, and $\tau\Gamma_{1,R}=3$.}\label{fig4}
\end{figure}

Now we turn to consider the non-Markovian regime, where $\tau$ is nonnegligible such that the phases $\phi_{1}=\phi_{1,0}+\tau\Delta$ and $\phi_{2}=\phi_{2,0}+\tau\Delta$ strongly depend on the detuning $\Delta$. Experimentally, this regime corresponds to the case of transmons coupled with surface acoustic waves~\cite{GuoSAW,SAWcase}, or the case where the separation $d$ between the two coupling points is large enough~\cite{longhiretard}. 

Owing to the strong $\Delta$-dependence of $\phi_{1}$ and $\phi_{2}$, one can expect in this case more complicated transmission spectra with staggered peaks and dips, as shown in Fig.~\ref{fig4}. Once again, we plot in Figs.~\figpanel{fig4}{a} and \figpanel{fig4}{b} the inelastic transmission rate $T_{2}$ of a left-incident photon versus the detuning $\Delta$ and the phase factor $\phi$ in two cases, i.e., (i) $\phi_{1,0}=\phi$, $\phi_{2,0}=0$ and (ii) $\phi_{1,0}=-\phi_{2,0}=\phi$, respectively, and plot in Figs.~\figpanel{fig4}{c} and \figpanel{fig4}{d} the corresponding two-dimensional profiles with some specific values of $\phi$. In both cases, the multiple dips with $T_{2}=0$ in the inelastic transmission spectra arise from the fact that perfect elastic transmission occurs whenever $\phi_{1}=\phi_{1,0}+\tau\Delta=(2m+1)\pi$ and/or $\phi_{2}=\phi_{2,0}+\tau\Delta=(2m+1)\pi$. Such non-Markovian characteristics can also be understood from the effective detuning shown in Appendix~\ref{appb}.

For the case of $\phi_{1,0}=\phi$ and $\phi_{2,0}=0$ as shown in Figs.~\figpanel{fig4}{a} and \figpanel{fig4}{c}, as $\phi$ increases from $0$, each conversion dip splits into two parts, with one of them moving toward the red-shift direction and the other one keeping still [their positions do not change with $\phi$ because $\phi_{2}=\tau\Delta=(2m+1)\pi$ is always satisfied there, which is the condition of perfect elastic transmissions and vanishing frequency conversions]. The moving dips can be understood from the fact that, the values of $\Delta$ that satisfy $\phi_{1}=(2m+1)\pi$ should change gradually with $\phi$. When $\phi=2m\pi$, the moving and static dips coincide because $\phi_{1}=\phi_{2}$ in this case. In view of this, one can control the positions of the conversion dips (or say, the positions of the elastic transmission peaks) flexibly by tuning the phase factor $\phi$ in this case. However, efficient frequency conversion of $T_{2}=1-T_{1}=1$ is unachievable for $\phi\rightarrow(2m+1)\pi$ in this case [see the green dot-dashed line in Fig.~\figpanel{fig4}{c} for instance]. This can be seen from the condition $\Delta^{2}+2[\Gamma_{1,R}^{2}-\Gamma_{1,R}^{2}\cos{(\phi_{1}+\phi_{2})}-\Gamma_{1,R}\Delta(\sin{\phi_{1}}+\sin{\phi_{2}})]=0$ of $T_{2}=1$ in the ideal chiral regime, which is directly obtained from Eq.~(\ref{eq11}). Clearly, this condition cannot be fulfilled if $\phi_{1}=\pi+\tau\Delta$ and $\phi_{2}=\tau\Delta$ as in Fig.~\figpanel{fig4}{a}. On the other hand, for the case of $\phi_{1,0}=-\phi_{2,0}=\phi$ as shown in Figs.~\figpanel{fig4}{b} and \figpanel{fig4}{d}, as $\phi$ increases from $0$, each dip splits into two parts which move toward opposite directions and merge with a new one at $\phi=\pi$. In this case, efficient frequency conversion can always be achieved and one can find again an ultra-narrow conversion window with $T_{2}=1$ for $\phi\rightarrow\pi$. 

\subsection{Comparison of the two regimes}

\begin{figure}[ptb]
\centering
\includegraphics[width=8.5 cm]{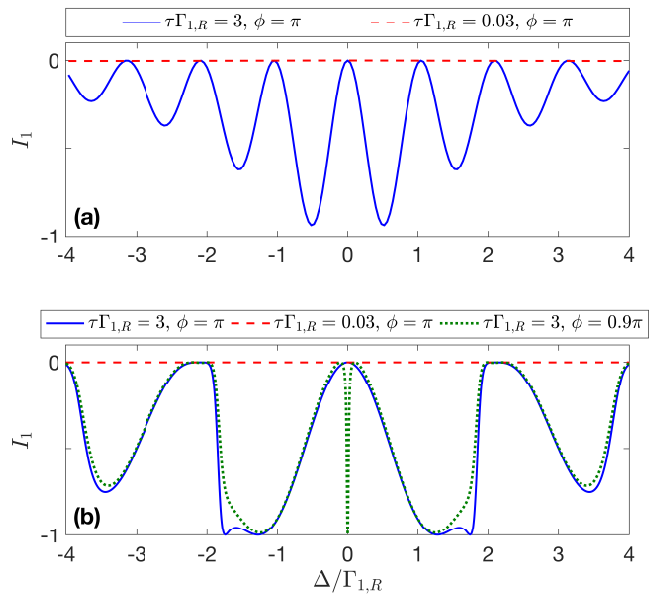}
\caption{Profiles of transmission contrast $I_{1}$ of the giant-atom model, with different values of $\tau$ and $\phi$ in the case of (a) $\phi_{1,0}=\phi$, $\phi_{2,0}=0$ and (b) $\phi_{1,0}=-\phi_{2,0}=\phi$. The Markovian and non-Markovian regimes correspond to the lines with $\tau\Gamma_{1,R}=0.03$ and $\tau\Gamma_{1,R}=3$, respectively. Here we assume $\Gamma_{1,R}=\Gamma_{2,R}$ and $\Gamma_{1,L}=\Gamma_{2,L}=0$.}\label{fig5}
\end{figure}

Finally, we plot in Fig.~\ref{fig5} the transmission contrast $I_{1}$ to compare the nonreciprocal scatterings of the giant-atom model in the Markovian and non-Markovian regimes. In particular, we focus on the case of $\phi\rightarrow(2m+1)\pi$ [$\phi_{1,0}=\phi$, $\phi_{2,0}=0$ in Fig.~\figpanel{fig5}{a} and $\phi_{1,0}=-\phi_{2,0}=\phi$ in Fig.~\figpanel{fig5}{b}], which shows significant difference for the two regimes. It can be seen that in the Markovian regime, the transmissions are always reciprocal because perfect elastic transmission takes place in both directions when $\phi_{1}=(2m+1)\pi$ and/or $\phi_{2}=(2m+1)\pi$. In the non-Markovian regime, however, $\phi_{1}$ and $\phi_{2}$ are sensitive to $\Delta$ such that nonreciprocal transmissions are allowed except for some specific values of $\Delta$. We would like to refer to such a phenomenon as \emph{non-Markovicity induced nonreciprocity}. 

For the case of $\phi_{1,0}=\phi$ and $\phi_{2,0}=0$, as shown in Fig.~\figpanel{fig5}{a}, the non-Markovicity induced nonreciprocity is not perfect (i.e., $|I_{1}|<1$) because efficient frequency conversions (vanishing elastic transmissions) are not allowed for a left-incident photon in this case [see again the green dot-dashed line in Fig.~\figpanel{fig4}{c}]. For the case of $\phi_{1,0}=-\phi_{2,0}=\phi$, as shown in Fig.~\figpanel{fig5}{b}, the nonreciprocal transmissions in the non-Markovian regime tend to be perfect whenever the condition of $T_{2}=1$, i.e., $\Delta^{2}+2[\Gamma_{1,R}^{2}-\Gamma_{1,R}^{2}\cos{(\phi_{1}+\phi_{2})}-\Gamma_{1,R}\Delta(\sin{\phi_{1}}+\sin{\phi_{2}})]=0$, is fulfilled. That is to say, in the ideal chiral case, perfect nonreciprocal elastic scatterings can be achieved if efficient frequency conversion is allowed. More importantly, as discussed above, the nonreciprocal window can be ultra-narrow in this case if $\phi\rightarrow(2m+1)\pi$, which enables precise filtering (or frequency conversion) in one direction but not in the other one.  

\subsection{Effect of different group velocities}\label{s3d}

\begin{figure}[ptb]
\centering
\includegraphics[width=8.5 cm]{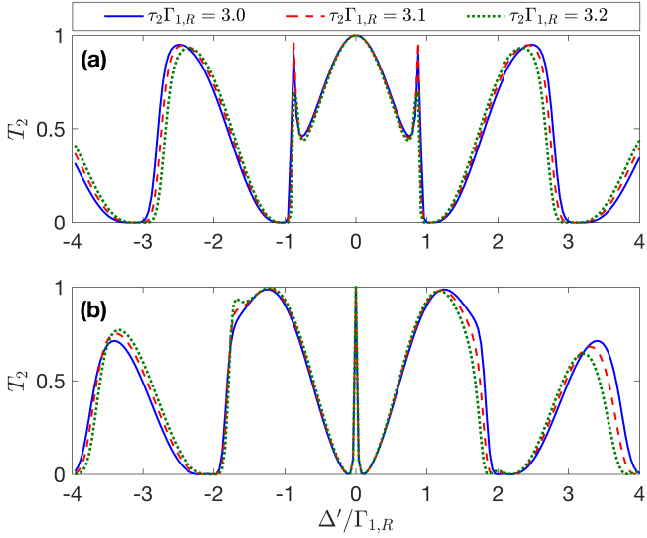}
\caption{Profiles of inelastic transmission rate $T_{2}$ of the giant-atom model, with different values of $\tau_{2}$ and (a) $\phi=0$ and (b) $\phi=0.9\pi$ in the case of $\phi_{1,0}'=-\phi_{2,0}'=\phi$. Here we assume $\Gamma_{1,R}=\Gamma_{2,R}$, $\Gamma_{1,L}=\Gamma_{2,L}=0$, and $\tau_{1}\Gamma_{1,R}=3$.}\label{fig6}
\end{figure}

Finally, we would like to demonstrate that the main results in this paper is almost unaffected even if the two lower states are not quasi-degenerate: in this case, the frequency difference $\omega_{f}$ between $|g\rangle$ and $|f\rangle$ is large enough so that the dispersion relation of the waveguide can be linearized with slightly different group velocities around the frequencies of the two lower states. Denoting the two group velocities by $v_{1}$ and $v_{2}$, respectively, the Hamiltonians $H_{\text{w}}$ and $H_{\text{int}}$ of the chiral giant-atom model can be rewritten as
\begin{equation}
\begin{split}
H_{\text{w}}&=\sum_{j=1,2}\int_{-\infty}^{+\infty}dx\Big[c_{j,L}^{\dag}(x)\Big(\omega_{j}+iv_{j}\frac{\partial}{\partial x}\Big)c_{j,L}(x)\\
&\quad\,+c_{j,R}^{\dag}(x)\Big(\omega_{j}-iv_{j}\frac{\partial}{\partial x}\Big)c_{j,R}(x)\Big],
\end{split}
\label{eq13}
\end{equation}
and
\begin{equation}
\begin{split}
H_{\text{int}}&=H_{\text{1,int}}+H_{\text{2,int}},\\
H_{\text{1,int}}&=\int_{-\infty}^{\infty}dx |g\rangle\langle e|\Big\{c_{1,R}^{\dag}(x)[g_{1,R}\delta(x)\\
&\quad\,+g_{2,R}\delta(x-d)]+c_{1,L}^{\dag}(x)[g_{1,L}\delta(x)\\
&\quad\,+g_{2,L}\delta(x-d)]\Big\}+\text{H.c.},\\
H_{\text{2,int}}&=\int_{-\infty}^{\infty}dx |f\rangle\langle e|\Big\{c_{2,R}^{\dag}(x)[\xi_{1,R}\delta(x)\\
&\quad\,+\xi_{2,R}\delta(x-d)]+c_{2,L}^{\dag}(x)[\xi_{1,L}\delta(x)\\
&\quad\,+\xi_{2,L}\delta(x-d)]\Big\}+\text{H.c.},
\end{split}
\label{eq14}
\end{equation}  
where $c_{j,R}^{\dag}(x)$ [$c_{j,L}^{\dag}(x)$] is the creation operator of the right-moving waveguide mode with group velocity $v_{j}$; $\omega_{1}$ ($\omega_{2}$) is the chosen frequency, around which the dispersion relation of the waveguide can be linearized as $\omega_{1}+v_{1}k$ ($\omega_{2}+v_{2}q$)~\cite{shen2009} and satisfies $\omega_{1}+v_{1}k=\omega_{2}+v_{2}q+\omega_{f}$ due to the energy conservation. Note that we do not assume $g_{j,R}=\xi_{j,R}$ and $g_{j,L}=\xi_{j,L}$ in this case. In the single-excitation subspace, the eigenstate can be rewritten as
\begin{equation}
\begin{split}
|\psi\rangle&=\int_{-\infty}^{+\infty}dx \Big\{[R_{g}(x)c_{1,R}^{\dag}(x)+L_{g}(x)c_{1,L}^{\dag}(x)]|0,g\rangle\\
&\quad\,+[R_{f}(x)c_{2,R}^{\dag}(x)+L_{f}(x)c_{2,L}^{\dag}(x)]|0,f\rangle\Big\}+u_{e}|0,e\rangle,
\end{split}
\label{eq15}
\end{equation}
with which one can analytically calculate the transmission coefficients with the same procedure above. In particular, with the assumption of $|g_{j,R}|^{2}/v_{1}=|\xi_{j,R}|^{2}/v_{2}=\Gamma_{j,R}$ and $|g_{j,L}|^{2}/v_{1}=|\xi_{j,L}|^{2}/v_{2}=\Gamma_{j,L}$, one can obtain identical transmission coefficients as those in Eq.~(\ref{eq11}), except that the detuning should be modified as $\Delta\rightarrow\Delta'=\omega_{1}-\omega_{e}+v_{1}k$. Moreover, the two different group velocities give rise to different propagation time for photons with wave vector $k$ and $q$ propagating between the two coupling points. Therefore the two phases are modified as $\phi_{1}=\phi_{1,0}'+\tau_{1}\Delta'=\phi_{1,0}'+d\Delta'/v_{1}$ and $\phi_{2}=\phi_{2,0}'+\tau_{2}\Delta'=\phi_{2,0}'+d\Delta'/v_{2}$, with $\phi_{1,0}'=(\omega_{e}-\omega_{1})\tau_{1}$ and $\phi_{2,0}'=(\omega_{e}-\omega_{f}-\omega_{2})\tau_{2}$ in this case. 

Now we assume that $v_{1}$ and $v_{2}$ are slightly different, and plot in Figs.~\figpanel{fig6}{a} and \figpanel{fig6}{b} the profiles of $T_{2}$ in the case of $\phi_{1,0}'=-\phi_{2,0}'=\phi$. In Fig.~\ref{fig6}, we focus on the non-Markovian regime where the difference between $\tau_{1}$ and $\tau_{2}$ introduces nonnegligible influence to the scattering behaviors. One can find that the single-photon scatterings of the giant-atom model are insensitive to the time difference, especially for $\Delta'\rightarrow0$. That is to say, the main results in this paper, such as the ultra-narrow scattering windows located at $\Delta'=0$, hold even if the two lower states $|g\rangle$ and $|f\rangle$ are not quasi-degenerate.     

\section{Conclusion \red{and outlook}}

In summary, we have generalized a chiral quantum optical model, where both the transitions of a $\Lambda$-type atom couple chirally to the waveguide field, to its giant-atom version by assuming that the chiral $\Lambda$-type atom interacts twice with the waveguide at two separated points. We have studied their nonreciprocal single-photon scatterings that are impossible for the case of a chiral two-level atom in the absence of intrinsic atomic dissipations. Compared with the small-atom case, the chiral giant-atom model exhibits intriguing interference effects that are closely related to the propagation time of photons between the two atom-waveguide coupling points, such as ultra-narrow scattering windows. In particular, we have explored the scattering behaviors of the chiral giant-atom model in both the Markovian and non-Markovian regimes, which are defined depending on whether the propagation time is negligible or not. It has been shown that the scattering behaviors are quite different in the two regimes, especially when either of the two atomic transitions is completely suppressed due to the destructive interferences: in this case, the scatterings are always reciprocal in the Markovian regime, while the non-Markovicity induced nonreciprocity with multiple nonreciprocal scattering windows can be observed in the non-Markovian regime. The results in this paper provide a deeper sight into the combination of chiral quantum optics and giant-atom physics, and have potential applications in, e.g., integrated photonics and quantum network engineering.   

We would like to point out that the investigations of multi-level giant atoms are still at the initial stage. More peculiar phenomena can be expected based on the marriage of multi-level giant atoms and various quantum effects. For instance, one can consider a $\Delta$-type giant atom, whose interference effect arising from its closed-loop level structure may bring more intriguing effects. Moreover, one can also extend some exotic effects for two-level giant atoms to their multi-level versions, such as oscillating bound states in the non-Markovian regime~\cite{oscillate} and decoherence-free interactions between giant atoms~\cite{decofree1,decofree2,decofree3}.  

\section*{Acknowledgments}

This work is supported by the National Natural Science Foundation of China (under Grants No. 11774024, No. 12074030, and No. U1930402) and the Science Challenge Project (Grant No. TZ2018003).

\appendix
\section{Transmission coefficients of a chiral two-level small atom}\label{app1}

For a chiral two-level small atom with the ground state $|g\rangle$ and the excited state $|e\rangle$, the Hamiltonian can be written as
\begin{equation}
\begin{split}
H'&=H_{\text{a}}'+H_{\text{w}}+H_{\text{int}}',\\
H_{\text{a}}'&=\omega_{e}|e\rangle\langle e|,\\
H_{\text{int}}'&=\int_{-\infty}^{\infty}dx \delta(x)[g_{R}'c_{R}^{\dag}(x)|g\rangle\langle e|\\
&\quad\,+g_{L}'c_{L}^{\dag}(x)|g\rangle\langle e|+\text{H.c.}],
\end{split}
\label{eqa1}
\end{equation}
where $g_{R}'$ ($g_{L}'$) is the coupling strength between the atom and the right-moving (left-moving) waveguide mode in this case; $H_{\text{w}}$ is identical with that in Eq.~(\ref{eq1}).

Once again, with the single-excitation state 
\begin{equation}
\begin{split}
|\phi\rangle&=w_{e}|0,e\rangle+\int_{-\infty}^{+\infty}dx[\phi_{R}(x)c_{R}^{\dag}(x)\\
&\quad\,+\phi_{L}(x)c_{L}^{\dag}(x)]|0,g\rangle,
\end{split}
\label{eqa2}
\end{equation}
where $\phi_{R}(x)$ [$\phi_{L}(x)$] denotes the wave function of the right-moving (left-moving) waveguide mode and $w_{e}$ denotes the probability amplitude of exciting the atom in this case, one can solve the stationary Schr\"{o}dinger equation and then obtain
\begin{equation}
\begin{split}
E\phi_{R}(x)&=\Big(\omega_{0}-iv_{g}\frac{\partial}{\partial x}\Big)\phi_{R}(x)+g_{R}'w_{e},\\
E\phi_{L}(x)&=\Big(\omega_{0}+iv_{g}\frac{\partial}{\partial x}\Big)\phi_{L}(x)+g_{L}'w_{e},\\
\delta w_{e}&=g_{R}'\phi_{R}(0)+g_{L}'\phi_{L}(0)
\end{split}
\label{eqa3}
\end{equation}
with $\delta=E-\omega_{e}=\omega_{0}+v_{g}k-\omega_{e}$. Assuming
\begin{equation}
\begin{split}
\phi_{R}(x)&=[\Theta(-x)+t\Theta(x)]e^{ikx},\\
\phi_{L}(x)&=r\Theta(-x)e^{-ikx}
\end{split}
\label{eqa4}
\end{equation}
for a left-incident photon and
\begin{equation}
\begin{split}
\phi_{R}(x)&=\tilde{r}\Theta(x)e^{ikx},\\
\phi_{L}(x)&=[\Theta(x)+\tilde{t}\Theta(-x)]e^{-ikx}
\end{split}
\label{eqa5}
\end{equation}
for a right-incident photon, the transmission coefficients can be readily solved as
\begin{equation}
\begin{split}
t&=\frac{2\delta+i(\Gamma_{L}'-\Gamma_{R}')}{2\delta+i(\Gamma_{L}'+\Gamma_{R}')},\\
\tilde{t}&=\frac{2\delta+i(\Gamma_{R}'-\Gamma_{L}')}{2\delta+i(\Gamma_{L}'+\Gamma_{R}')}.
\end{split}
\label{eqa6}
\end{equation}

Clearly, $|t|^{2}\equiv|\tilde{t}|^{2}$ in this case, implying that the single-photon scatterings here are always reciprocal even if the atom-waveguide couplings are chiral. Nevertheless, if the intrinsic decay rate $\kappa$ of the atom is nonnegligible (the dissipation from the atom to the environment outside the waveguide is taken into account), $\delta=E-\omega_{e}+i\kappa$ becomes complex and thus nonreciprocal scatterings ($|t|^{2}\neq|\tilde{t}|^{2}$) can be achieved. On the other hand, such a chiral two-level atom is in fact equivalent to a giant atom with nontrivial phase difference between its different atom-waveguide coupling channels~\cite{Zoller1,Zoller2}, which mimics an Aharonov-Bohm cage that cannot exhibit nonreciprocal transmissions without non-Hermitian on-site potentials~\cite{sz1,sz2,sz3}. 

\section{Effective detuning and decay rate of the chiral giant-atom model}\label{appb}

In this section, we provide an analytical way to understand the unltra-narrow scattering windows in the giant-atom case. For the chiral giant-atom model considered in this paper, both the transition frequencies and the total radiative decay rate of the atom are modified due to the giant-atom interference effects~\cite{Lamb,DLlambda}. The effective detuning and decay rate can be given by the real and imaginary parts of the denominator in Eq.~(\ref{eq11}), respectively, i.e.,
\begin{equation}
\begin{split}
\Delta_{\text{eff}}&=\Delta-(\Gamma_{12,R}+\Gamma_{12,L})(\sin{\phi_{1}}+\sin{\phi_{2}}),\\
\Gamma_{\text{eff}}&=\Gamma_{1,R}+\Gamma_{2,R}+\Gamma_{1,L}+\Gamma_{2,L}\\
&\quad\,+(\Gamma_{12,R}+\Gamma_{12,L})(\cos{\phi_{1}}+\cos{\phi_{2}}).
\end{split}
\label{eqb1}
\end{equation}
For the ideal chiral case of $\Gamma_{1,R}=\Gamma_{2,R}=\Gamma_{R}$ and $\Gamma_{1,L}=\Gamma_{2,L}=0$, Eq.~(\ref{eqb1}) can be simplified as
\begin{equation}
\begin{split}
\Delta_{\text{eff}}&=\Delta-\Gamma_{R}[\sin{\phi}\cos{\tau\Delta}+\sin{\tau\Delta}(1+\cos{\phi})],\\
\Gamma_{\text{eff}}&=\Gamma_{R}[2+\cos{\tau\Delta}(1+\cos{\phi})-\sin{\phi}\sin{\tau\Delta}]
\end{split}
\label{eqb2}
\end{equation}
if $\phi_{1,0}=\phi$ and $\phi_{2,0}=0$, and simplified as
\begin{equation}
\begin{split}
\Delta_{\text{eff}}&=\Delta-2\Gamma_{R}\cos{\phi}\sin{\tau\Delta},\\
\Gamma_{\text{eff}}&=2\Gamma_{R}(1+\cos{\phi}\cos{\tau\Delta})
\end{split}
\label{eqb3}
\end{equation}
if $\phi_{1,0}=-\phi_{2,0}=\phi$. 

Clearly, in the case of Eq.~(\ref{eqb2}), the effective detuning is always $\phi$ dependent in both the Markovian and non-Markovian regimes, as shown in Figs.~\figpanel{fig3}{a} and \figpanel{fig4}{a}. In the case of Eq.~(\ref{eqb3}), the effective detuning does not depend on $\phi$ in the Markovian regime where $\tau\Delta\rightarrow0$ always holds, as shown in Fig.~\figpanel{fig3}{b}, whereas in the non-Markovian regime only the position of the central scattering window (located at $\Delta=0$) does not depend on $\phi$ due to $\tau\Delta=0$, as shown in Fig.~\figpanel{fig4}{b}.

On the other hand, one can find from Eq.~(\ref{eqb3}) that the width of the central scattering window that is proportional to $\Gamma_{\text{eff}}(\Delta=0)$ tends to zero as $\phi$ increases from $2m\pi$ to $(2m+1)\pi$. Although in the case of Eq.~(\ref{eqb2}), $\Gamma_{\text{eff}}(\Delta=0)$ decreases as well when $\phi$ increases in the range of $[2m\pi,(2m+1)\pi]$, there is a nonzero lower bound $\text{min}[\Gamma_{\text{eff}}(\Delta=0)]=2\Gamma_{1,R}$, with which the unltra-narrow scattering windows are not available.   

Finally, we point out that the difference between the scattering spectra in the Markovian and non-Markovian regimes can also be understood from Eqs.~(\ref{eqb2}) and (\ref{eqb3}). For $\tau\Delta\rightarrow0$ in the Markovian regime, the Lamb shifts $\Delta_{\text{eff}}-\Delta$ induced by the giant-atom effects are constant or depend only on $\phi$. In the non-Markovian regime, however, the Lamb shifts become $\Delta$ dependent as well such that the spectra exhibit multiple windows~\cite{GuoSAW,SAWcase}.

\end{document}